# Chromatin Loops as Allosteric Modulators of Enhancer-Promoter Interactions

Boryana Doyle[1,2,3], Geoffrey Fudenberg[4], Maxim Imakaev[3], and Leonid A. Mirny[3,4,5]

[1] Program for Research in Mathematics, Engineering and Science for High School Students, PRIMES, Massachusetts Institute of Technology
[2] MIT Undergraduate Research Opportunities Program (UROP)
[3] Department of Physics, Massachusetts Institute of Technology
[4] Graduate Program in Biophysics, Harvard University
[5] Institute for Medical Engineering and Science, Massachusetts Institute of Technology

*Corresponding Author: Leonid A. Mirny, E25-526, MIT, 77 Mass Ave. Cambridge MA 02139, 617-452-4862, email: leonid@mit.edu

## Abstract

The classic model of eukaryotic gene expression requires direct spatial contact between a distal enhancer and a proximal promoter. Recent Chromosome Conformation Capture (3C) studies show that enhancers and promoters are embedded in a complex network of looping interactions. Here we use a polymer model of chromatin fiber to investigate whether, and to what extent, looping interactions between elements in the vicinity of an enhancer-promoter pair can influence their contact frequency. Our equilibrium polymer simulations show that a chromatin loop, formed by elements flanking either an enhancer or a promoter, suppresses enhancer-promoter interactions, working as an insulator. A loop formed by elements located in the region between an enhancer and a promoter, on the contrary, facilitates their interactions. We find that different mechanisms underlie insulation and facilitation; insulation occurs due to steric exclusion by the loop, and is a global effect, while facilitation occurs due to an effective shortening of the enhancer-promoter genomic distance, and is a local effect. Consistently, we find that these effects manifest quite differently for *in silico* 3C and microscopy. Our results show that looping interactions that do not directly involve an enhancer-promoter pair can nevertheless significantly modulate their interactions. This phenomenon is analogous to allosteric regulation in proteins, where a conformational change triggered by binding of a regulatory molecule to one site affects the state of another site.

## Author Summary

In eukaryotes, enhancers directly contact promoters over large genomic distances to regulate gene expression. Characterizing the principles underlying these long-range enhancer-promoter contacts is crucial for a full understanding of gene expression. Recent experimental mapping of chromosomal interactions by the Hi-C method shows an intricate network of local looping interactions surrounding enhancers and promoters. We model a region of chromatin fiber as a long polymer and study how the formation of loops between certain regulatory elements can insulate or facilitate enhancer-promoter interactions. We find 2-5 fold insulation or facilitation, depending on the location of looping elements relative to an enhancer-promoter pair. These effects originate from the polymer nature of chromatin, without requiring additional mechanisms beyond the formation of a chromatin loop. Our findings suggest that loop-mediated gene regulation by elements in the vicinity of an enhancer-promoter pair can be understood as an allosteric effect. This highlights the complex effects that local chromatin organization can have on gene regulation.

# Introduction

Distal enhancer elements in higher eukaryotes are essential for regulating gene expression [1-4]. In conjunction with transcription factor binding and nucleosome modifications, the classic model of enhancer function requires the direct spatial contact between enhancers and their target promoters (Figure **1A**) [1-4]. Recent studies have started to reveal the complexity of the enhancer-promoter (E-P) interaction network, where each enhancer can influence multiple promoters, and each promoter may be influenced by multiple enhancers [5-8]. In addition, gene expression and E-P interactions occur within higher-order three-dimensional chromatin organization, which is characterized by an intricate network of interactions at multiple scales. For example, below 1Mb, chromatin is organized into continuous 500-900kb regions of enriched contact frequency called topologically associated domains (TADs) [9,10]. TADs were found to be about 90% cell-type independent (2763/3000 conserved boundaries between two assayed cell types [9]). Within TADs, additional cell-type specific looping interactions are formed [6,11,12]. These observations raise an important question; namely, how can E-P contacts be affected by looping interactions between other regulatory elements in their genomic neighborhood?

Two models for how proximal looping interactions can modulate E-P contacts have been proposed: the decoy model and the topological model (experiments [13-15], reviewed in [16-19]). The decoy model suggests that insulating elements directly interact with the enhancer, sequestering it from the promoter, and thereby directly hinder E-P interactions. The topological model proposes that two regulatory elements in the vicinity of the enhancer and promoter can interact with each other to form a chromatin loop; this, in turn, affects E-P contacts.

Evidence supporting the topological model includes experiments in multiple organisms (Figure **1B-D**) [20-23]. For example, Kyrchanova et al. [21] recently observed that a single Drosophila *gypsy* element placed between an enhancer and a promoter did not change their interactions; however, introducing two *gypsy* elements changed E-P interactions depending on *gypsy* position and orientation. The authors explain these observations by *gypsy-gypsy* looping interactions. We note that while this *gypsy* element consisted of twelve repeated copies of the Su(Hw) binding site, elements with fewer sites are sufficient for insulation [24,25]. The regulatory effects of loops may also be relevant at larger genomic distances; in mice, a regulatory element with multiple larger (25kb and 55kb) loops was suggested to control multiple E-P contacts at the H19 locus [22]. Analogously, loops between insulating elements were suggested to modulate the activity of silencing elements [23].

It remains unclear whether, and to what extent, the looping interactions between other regulatory elements can mediate E-P contacts. When these looping interactions do not directly involve the E-P pair, their effect is reminiscent of allosteric regulation in proteins [26,27], where binding of a regulator molecule to one site changes the state of another site or the whole protein. A classic example is the binding of allolactose to the lac repressor at the regulatory domain. While allosteric interactions in proteins are mediated by the protein structure, we propose that interactions between genomic sites could be mediated by local changes in the conformational ensemble of the chromatin fiber.

Polymer simulations provide an ideal testing ground to investigate the allosteric effects of a loop on E-P contacts; many loci can be probed simultaneously at high resolution, and more complicated looping arrangements can be systematically characterized. Previously, Mukhopadhyay et al. [28] used polymer simulations to demonstrate that the topological model of insulation applies to an unconfined system of two fused chromatin rings; namely, two loci within the same ring interact more frequently than loci in different rings. We extend this line of inquiry by asking whether forming loops may affect interactions at scales exceeding the loop size, e.g. interactions of a loop with the rest of the chromosome or between loci in the vicinity of the loop.

Here we use polymer models to study how 15-60kb chromatin loops can influence E-P contacts. We note that 3C-based studies have only begun to provide unbiased data at sufficient resolution to build polymer models of a particular locus [29], and the fine structure of the chromatin fiber *in vivo* remains largely unknown [30,31]. Thus, for generality, we model chromatin as a long homogeneous flexible fiber with only a few additional looping interactions between specific elements, as described below. Synthesizing results from the literature, we primarily focused our simulation analysis on two important arrangements of the loop-forming elements relative to an E-P pair: (1) an enhancer is flanked by loop-forming elements, while a promoter is beyond the loop (Figure **1E**); and (2) both loop-forming elements are located in the genomic region between an enhancer and a promoter (Figure **1F**).

We find that loops can significantly insulate or facilitate the frequency of E-P interactions, depending on the loop location relative to the E-P pair. We consider a variety of situations and parameters, including: E-P genomic distance, stiffness of the chromatin fiber, size of the loop, topological constraints on the chromatin fiber (i.e. topoisomerase II activity), chromatin density, the number of looping elements, and excluded volume interactions. We find that different mechanisms underlie insulation and facilitation; insulation occurs due to steric exclusion by the loop, while facilitation occurs due to an effective shortening of the E-P genomic distance. We additionally consider how insulation and facilitation would be observed in microscopy studies and find substantial differences from how they would manifest in 3C-based studies. Taken together, our results suggest that due to its polymer nature, chromatin allows for interactions to be mediated in an allosteric manner, i.e. formation of a contact between two sites can insulate or facilitate interactions between other loci in the vicinity.

## Results

**Model and analysis of simulations**

Using equilibrium simulations of a confined polymer chain, we study how chromatin loops affect E-P contact frequency in their vicinity. We model chromatin as a semi-flexible polymer fiber with excluded volume; the fiber consists of 15nm diameter monomers, each representing three nucleosomes or 500bp, with a persistence length of 3 monomers (Figure **1G,** Methods) [32]. Unless otherwise noted, we allow occasional chromatin fiber crossing by setting a finite energy cost (using a truncated repulsive potential) for two monomers to occupy the same volume, which accounts for topoisomerase II (topo-II) activity (see Methods). Thus, two regions of the chain can spontaneously cross through each other with a probability controlled by the energy penalty of co-occupancy. To account for the dense arrangement of chromatin within the nucleus, we confine the chromatin fiber to impose a 2% volume density. We later vary volume density from 1% to 20% (see below), which is consistent with current estimates of chromatin volume density in the interphase nuclei of higher eukaryotes [33]. Since the flexibility of the chromatin fiber *in vivo* is incompletely characterized, we varied flexibility in our simulations and found quantitatively similar results (see below).

For each set of conditions and looping interactions, we performed Langevin dynamics simulations using OpenMM [34] (see Methods and Video S**1**) and sampled conformations from the resulting equilibrium ensemble; these conformations were subsequently analyzed to compute contact frequencies (see below, and Methods). To investigate the effects of a chromatin loop on a larger region of chromatin, we model a loop by forming an irreversible bond between a pair of monomers and allowing the whole polymer to equilibrate (Figure **1E**, **1F**, see Methods). We considered loops of sizes $L$=15kb, 30kb, and 60kb, and a 2.5kb loop with a more flexible chromatin fiber (see Discussion), in a proportionally sized genomic region of length 33*$L$, i.e. 1Mb for a 30kb loop (Figure **S1**). Our polymer model contains no additional sequence-specific details, and thus generally addresses how E-P interactions are altered in the vicinity of a loop. The model remains agnostic to the chromatin organization at larger genomic scales, assuming that the simulated region is contained within a single TAD [35].

For each set of parameter values and loops, we generate an equilibrium ensemble of conformations and compute the contact frequency between loci (monomers) in this ensemble (Figure **2A**, Table **S1** for parameter values). We display pairwise contact frequencies using heatmaps (Figure **2B**), as typical for Hi-C and 5C experiments. Our simulated heatmaps are characterized by two features: (i) a decay of contact frequency as a function of increasing genomic distance, and (ii) an off-diagonal interaction between the loop bases. The first feature follows from the polymer connectivity of the simulated chromatin fiber. The second feature alters the typical decline in the contact frequency and is of primary interest in this study.

For a given position of the enhancer and the promoter, we can compute the *contact frequency ratio* as the contact frequency in the model with a loop, divided by the contact frequency for an otherwise equivalent model without a loop. Contact frequency ratios below 1 indicate insulation, whereas ratios above 1 correspond to facilitation. Unless noted otherwise, we report contact frequency ratios for a 30kb loop and a 50kb E-P genomic distance. We note that each simulation contains information regarding every possible position of the enhancer and the promoter. From this, we can compute contact frequency ratios as a function of E-P distance and location. Below we examine how the loop length and the E-P spacing affect observed phenomena.

**Chromatin loops can insulate or facilitate enhancer-promoter interactions**

We used the simulated heatmaps of pairwise contact frequency to investigate the two important arrangements of the E-P pair and the loop from the literature (Figure **1**).

The first arrangement involves a chromatin loop formed by elements flanking an enhancer, such that the enhancer is located within the chromatin loop and the promoter is located outside of the loop (Figure **2C**). Since the enhancer and promoter are equivalent in our polymer model, this scenario also describes a promoter flanked by a pair of loop-forming elements and an enhancer located outside of the loop. Simulations show that for 50kb E-P spacing, formation of such a

30kb loop leads to a ~35% reduction in E-P contacts, serving as an insulator (contact frequency ratio of 0.64, Figure **2D**). Below we refer to this arrangement as insulation.

The second arrangement constitutes a chromatin loop located in the genomic region between the enhancer and promoter, i.e. both loop-forming elements are located between the enhancer and promoter (Figure **2C**). Formation of such a loop facilitates E-P interactions by increasing their contact frequency by more than 4-fold (contact frequency ratio of 4.15, Figure **2D**).

Next we examined how E-P spacing affects the magnitude of loop-induced insulation or facilitation. Interestingly, the two effects behave differently; while facilitation diminishes with E-P genomic distance, insulation appears to be independent of distance (Figure **3A**). These results reveal an important difference between loop-induced facilitation and insulation: facilitation is a local phenomenon, and insulation is a global effect.

To better understand insulation, we varied the position of the enhancer within the loop. We found that insulation is weaker when the enhancer is placed in the middle of the loop (0.75 contact frequency ratio), and strengthens as the enhancer approaches the base of the loop (0.49 contact frequency ratio, Figure **3B**). We note that an extreme case of topological-model insulation is in fact similar to decoy-model insulation, which occurs when the enhancer is placed at the base of the loop. In this scenario, we observe stronger insulation because the enhancer is permanently interacting with the other loop base, sterically hindering interactions between the enhancer and all other loci. This can be seen as dark stripes at the positions of the loop base monomers on the heatmap; the profile of interactions of the loop base with the rest of the fiber is detailed in Figure **S2**. Below we identify and discuss mechanisms underlying insulation and facilitation.

**Chromatin fiber flexibility, topological constraints, and overall density do not underlie insulation or facilitation**

To test the generality of insulation and facilitation, we varied several biologically relevant and physical characteristics of our model, many of which have not been fully characterized *in vivo*.

First, we investigated the importance of chromatin fiber flexibility by simulating chromatin fibers with different persistence lengths. We found that fiber flexibility does not significantly affect insulation or facilitation (Figure **S3**). This is consistent with the fact that both phenomena are observed at distances much larger than the persistence length, and thus in our simulations do not emerge solely due to fiber stiffness. As such, cartoons with rigid, stiff loops should in many cases be understood as schematics [36]; renderings of three-dimensional chromatin loops from our models are shown in Figure **1E**, **1F**, and **2A**. We also note that for simulations with larger and smaller loop sizes, the main qualitative features of the heatmap remain the same (Figure **S1**).

Next, we studied the effect of topological constraints, as they have been suggested to play an important role in chromosome organization [33,37,38]. To investigate this, we performed simulations both with and without allowing two regions of the chromatin fiber to cross, which may respectively correspond to cells with active and inactive topo-II. We found that insulation and facilitation are observed irrespective of the topological constraints (Figure **S4A**). We note that the terms topological model, topologically-associated domains (TADs), and topological constraints all refer to distinct, and likely unrelated, concepts. In particular, our results demonstrate that the topological model of insulation is independent from topological constraints on the chromatin fiber. Additionally, topological constraints are distinct from other topological effects such as supercoiling of the chromatin fiber [39]; supercoiling can lead to significant insulation and facilitation [39], but may be more relevant for bacterial chromosome organization [40].

Third, we assessed the influence of chromatin density on insulation and facilitation. In particular, active and inactive chromatin environments are known to have respectively lower and higher densities [33,41]. We performed simulations at densities ranging from low (1%) to high (20%) volume density (Figure **S4B**). We found that while both insulation and facilitation remain qualitatively present at all densities, they are roughly twice as strong at 1% vs. 20% density. This finding indicates that low density, as found in active chromatin, is important for the magnitude of both possible regulatory effects.

Together, these variations in our model suggest that insulation and facilitation exist across a range of biologically relevant parameter values. However, they indicate that insulation and facilitation do not mechanistically follow from the fiber stiffness, topological constraints, or overall density.

**Fundamental properties of polymers underlie insulation and facilitation**

To understand the mechanisms of insulation and facilitation, we performed simulations of a phantom polymer chain, which lacks excluded volume interactions (Figure **S5**). Remarkably, elimination of excluded volume interactions completely abolishes the insulation effect. In contrast, the degree of facilitation remains largely unaffected by the elimination of excluded volume (reduced from 4.15 to 3.20). We note that phantom chain simulations do not adequately describe chromosomes, but nevertheless can provide useful insights into polymer behavior; here, they demonstrate how steric exclusion by a loop can give rise to insulation.

The loss of insulation in simulations without excluded volume interactions led us to investigate the spatial relationship between the loop and the rest of the polymer fiber. We found that the spatial density of monomers from other regions of the fiber is depleted near the loop, i.e. the loop sterically excludes interactions with the rest of the polymer (Figure **4A**). Interestingly, regions immediately outside the loop are also sterically excluded by the loop; we find 20-50% insulation for regions up to 6kb away from the loop (Figure **S6A**).

Facilitation does not depend on excluded volume interactions, but depends on E-P distance. Therefore we considered how facilitation might arise from an effectively shortened E-P distance imposed by the intervening loop; in particular, we compared contact frequency ratios for the facilitation arrangement and for a simulation without a loop but at a 30kb smaller genomic distance (i.e. shortened by the loop size). Indeed, we see that these are in almost complete agreement, demonstrating that facilitation results from the effectively shortened genomic distance (Figure **4B**).

To get further insight into the mechanisms of insulation and facilitation, we performed *in silico* fluorescence *in situ* hybridization (FISH) by calculating the distribution of E-P spatial distances across many conformations (Figure **4C**). To consistently compare insulation and facilitation, we considered them at an E-P distance of 90kb, where both effects have approximately the same fold change (contact frequency ratios of 0.75 and 1.3). For insulation, we observe only a small shift in the overall distribution of E-P spatial distances (mean increased by 3%). This confirms that insulation occurs not because the E-P pair is much further away on average, but due to steric exclusion of the promoter by the loop engulfing the enhancer. For facilitation, however, the distribution of E-P spatial distances shifted more strongly (mean decreased by 9%). These results highlight that differences in contact frequency are not always proportionally reflected in differences in mean spatial distances. Moreover, our results show that both effects could be hard to detect by microscopy, but facilitation would be more evident than insulation.

Together, these results provide evidence for the mechanisms underlying insulation and facilitation. For insulation, regions within the loop are sterically excluded from making contacts with the rest of the polymer fiber. For facilitation, the E-P pair has an effectively shorter genomic distance.

**Intra-loop interactions and two-loop models**

The analyses above focused on understanding how a single loop affects E-P contact frequency at genomic distances exceeding the loop size. For E-P genomic distances less than the loop size, both elements can be positioned within the loop. In this case, we found that interactions are facilitated, consistent with previous results [28]. However, the degree of facilitation depends on the relative position of the elements in the loop (Figure **S6B**). Placing the enhancer at one loop base and the promoter at the other can greatly facilitate their interaction frequency. On the contrary, with the enhancer at the loop base and the promoter in the enhancer-proximal portion of the loop, the facilitation effect may disappear, likely due to the superposition of intra-loop facilitation with the insulating properties of the loop bases (Figure **S2B**).

Many enhancers, promoters, and loop-forming elements can be present in a given genomic region, opening the possibility for more complicated scenarios of insulation and facilitation. Towards this end, we performed simulations where two consecutive loops were formed. We observed qualitatively similar insulation and facilitation in the two-loop case, for the two arrangements similar to those we initially focused on (Figure **S7**). Within this two-loop element, the average contact frequency between loci within one loop is higher than the average contact frequency between loci from different loops. In this sense, the two loops are insulated from each other as well as from the rest of the fiber. This is consistent with simulations of an isolated system of two fused rings [28]. Moreover, these results show that the concept of steric exclusion, which underlies insulation for a single loop, applies to the two-loop case as well. In particular, each loop in the two-loop model sterically excludes the other, as well as the rest of the chromatin fiber.

# Discussion

Using a polymer model of chromatin, we found that a single loop in the vicinity of an E-P pair can either insulate or facilitate their interactions. These effects have a considerable magnitude, with about 2-fold insulation and 3-5 fold facilitation of E-P contact frequency, which is comparable to generally observed changes in gene expression [42].

Collectively, experiments have observed that different local arrangements of regulatory elements can lead to complex patterns of gene expression. For example, one insulating element between an E-P pair can decrease gene expression, yet two elements between the same E-P pair do not [15,16,18,19]; it was hypothesized that the two elements cancel each other out by forming a loop. Our model shows that this loop would in fact facilitate E-P interactions. Additionally, our model predicts that if the second element were placed outside the E-P pair, the resulting loop would indeed insulate E-P interactions. Note that since the exact quantitative relationship between E-P contact frequency and gene expression or phenotype remains largely unknown, we focus on qualitative comparisons between our model and these experimental studies.

The indirect modulation of E-P contacts by chromatin loops is often referred to as the topological model [16,17,19], a term used rather vaguely. Studies that consider the topological model often assume a particular mechanism whereby the loop alters E-P interactions. Specific mechanisms include: sliding along DNA [43], lamina attachments [19,44], and inter-nucleosome interactions [17]. Our simulations show that the formation of the loop itself can insulate or facilitate E-P interactions, due to the polymer nature of chromatin, independent of specific molecular mechanisms.

These effects can be best understood in terms of allosteric regulation. In particular, interactions that are responsible for the formation of the loop do not necessarily directly prevent or form E-P contacts. Instead, they steer the conformational ensemble of the chromatin fiber toward or away from conformations where an enhancer and a promoter are in contact. This mechanism of action is analogous to classical allosteric regulation in proteins [26], and particularly to disordered proteins, where binding of an allosteric substrate changes the protein conformation, which in turn alters the structure of a distant active site [45]. We note that the concept of allostery has also been useful for understanding other systems, including nucleosome-mediated transcription factor binding cooperativity [46,47].

The polymer mechanisms underlying insulation and facilitation arise due to two different effects. Facilitation results from the effectively shortened E-P genomic distance due to the loop and is likely robust to the molecular details of the chromatin fiber. Insulation arises due to excluded volume interactions and steric exclusion by the loop, and thus depends on the chromatin-chromatin affinity. In polymer physics terms, insulation would require good solvent conditions, which are likely satisfied in active decondensed chromatin. We further note that altering chromatin-chromatin affinity may allow further modulation of insulation strength, for example through chromatin modifications that change the net charge of nucleosome tails.

In specific biological systems, the detailed structure and flexibility of the chromatin fiber may become relevant. When the chromatin loop size approaches several persistence lengths, the loop could become very rigid. Consequentially, its effects may depend on the molecular details of the loop-forming elements, including their orientation as observed in a recent study [21]. However, the insulation and facilitation we observe may still manifest with similar strength at smaller genomic distances for more flexible or loosely packed chromatin fiber; changing these parameters causes small loops to behave similarly to larger loops (Figure **S1D**). We note that many processes may locally increase chromatin flexibility, either uniformly or through the formation of kinks [38], including the loss/unwrapping of nucleosomes [48]. Finally, the fine details of loop formation may be very important when the E-P pair is within the loop or near the loop bases, as we observed large variations in facilitation from subtle differences in E-P position in these cases.

Reconciling views of chromosome organization from 3C-based and microscopy studies remains an important challenge [49]. In our simulations, we found that changes in contact frequency are not always accompanied by equal changes in the mean spatial distance between two loci. In particular, insulation changes the distribution of spatial distances at small values, while having little effect on the mean. Changes in the spatial distribution at small distances could be difficult to detect experimentally and would require many cells to be assayed. Our results also suggest that integrating 3C-based and microscopy data can provide mechanistic insights.

Another important aspect of *in vivo* networks of local looping interactions is that they may be both dynamic over the course of the cell cycle [32] and different between cells [50]. Our results for insulation and facilitation by fixed loops, where the bases of the loop are always connected, remain relevant for dynamic loops while they are present. Roughly speaking, the effect on insulation or facilitation for a given loop is proportional to its frequency of occurrence in a cell.

Given the complexity of the local looping network, it is likely that there are multiple dynamic loops in the vicinity of the enhancer and promoter. While we studied the permanent single and double loop systems, our results provide intuition even to these more complicated systems. For instance, the global nature of insulation implies it can hinder interactions between enhancers and any number of promoters. Conversely, facilitation is local and thus specific to the regions that directly flank the loop. Together, our results highlight the complex and non-local grammar of regulatory elements surrounding enhancers and promoters. In conjunction with emerging biological data, future simulations will provide additional insight into the consequences of chromatin's polymer nature for allosteric modulation E-P interactions.

## Methods
**Polymer model**
*Model overview*: We modeled chromatin as a fiber of monomers connected by harmonic bonds. Unless noted, each spherical monomer had a diameter of 15 nm and represented 500bp, or approximately three nucleosomes. A permanent loop was formed by connecting two monomers with a harmonic bond of the same strength as the bonds between all consecutive monomers. This permanently brings the loop bases into contact. Two such loops were formed in the two-loop simulations. A three-point interaction force was used to impose a bending energy and account for the rigidity of the fiber. To model volume interactions, monomers interacted via a shifted Lennard-Jones potential, which is a computationally efficient purely-repulsive potential. Unless noted otherwise, the Lennard-Jones potential was truncated at $U=3kT$ as specified below to allow occasional fiber crossing. Simulated polymers were confined to a sphere at a given density and initialized from an unentangled polymer conformation.

Polymer models were simulated with OpenMM, a high-performance GPU-assisted molecular dynamics software (https://simtk.org/home/openmm). We used an in-house *openmm-polymer* library to efficiently set up polymer simulations with OpenMM, and to analyze simulation results. *openmm-polymer* is publicly available on the Bitbucket online repository: http://bitbucket.org/mirnylab/openmm-polymer. Scripts used to perform simulations, build contact maps, and calculate insulation/facilitation are available in the "examples" folder of the *openmm-polymer* library; those scripts can be modified to incorporate any arrangement of loops and calculate facilitation or insulation for any parameter values.

Simulations were characterized by 4 parameters: loop size, number of loops, fiber stiffness, and system density. Total polymer length was always chosen to be approximately 33 * loop size. The initial conformation for all simulations was an unentangled polymer ring. Simulations for a phantom chain were performed by switching off inter-monomer Lennard-Jones interactions. Choice of parameters for various models is summarized in Table **S1**.

**Forces and Langevin Dynamics simulations**
Adjacent monomers were connected by harmonic bonds with a potential $U = 25*(r - 1)^2$ (here and below, energy is in units of kT). The stiffness of the fiber was modeled by a three point interaction term, with the potential $U = k*(1-\cos(\alpha))$, where α is an angle between neighboring bonds, and *k* is a parameter controlling stiffness. A value of k=3 was used for most simulations; k=2 and k=4 were used for simulations with lower and higher stiffness; k=2 was used for simulations with the smallest (10 monomer) loop.

Neighboring monomers interacted via a shifted Lennard-Jones (LJ) repulsive potential $U = 4 * (1/r^{12} - 1/r^6) + 1$, for $r<2^{1/6}$; U=0 for $r > 2^{1/6}$ (for details see [32]). To account for the activity of type-II topoisomerase, we allowed fiber crossing by truncating the shifted LJ potential at an energy of $E_{cutoff}$ = 3 kT. For energy U more than 0.5 * $E_{cutoff}$, the LJ potential was modified as: $U_{softened}$ = 0.5 * $E_{cutoff}$ * (1 + tanh(2*U/$E_{cutoff}$ - 1)). To avoid numerical instabilities, the interaction radius r was truncated at r = 0.3 via: $r_{truncated} = (r^{10} + 0.3^{10})^{1/10}$, which introduced a negligible shift in the final softened potential. We note that our simulations were performed at thermodynamic equilibrium, and thus the rate at which the fibers were allowed to pass through each other does not influence the equilibrium properties of the system; the only relevant factor is whether the system is allowed to change its topological state or is "locked" to the unknotted topological state. We explored both scenarios in our simulations. Spherical confinement was realized as a potential linearly increasing at a rate of k=5kT/mon when the radius was larger than the confinement radius.

We simulated our model using Langevin Dynamics, performing 80,000 blocks of 3000 time steps (240,000,000 time steps total). For the fiber lengths considered here, polymer simulations reached equilibrium in less than 1000 blocks; this was confirmed by observing that monomer displacement saturates at about 500 blocks. Polymer conformations starting with block 1000 were used for our analysis. We note that this study focuses on equilibrium aspects of chromatin loops and that simulated time is not specifically matched to the time-scale of E-P interactions *in*

*vivo*; chromatin loop dynamics are beyond the scope of this study.

An Andersen thermostat was used to keep the kinetic energy of the system from diverging. The time step was then chosen to ensure conservation of kinetic energy and lack of fiber crossing with the non-truncated Lennard-Jones potential. The absence of fiber crossing in this case was confirmed by the conservation of Alexander's polynomial for a 50000-monomer ring simulated at a high density of 0.85 for 1,000,000,000 time steps.

**Initialization and starting conformations**
Since our simulations were performed at thermodynamic equilibrium, the starting conformation does not affect properties of the resulting heatmap; for simulations with fixed topology (i.e. no fiber crossing), only the topological state of the starting conformation is relevant. For simulations with or without fiber crossings, we initialized our simulations from an un-entangled polymer state created as described below. We started with a 4-monomer ring on a cubic lattice. We then chose one bond at random, and tried to extend the polymer at this location by two monomers, by making a bond into a kink. To do this, we considered another bond, obtained by shifting this bond by one in a random direction perpendicular to the bond (choosing one out of 4 possible directions). If both locations of the shifted bond were free, the polymer was extended to incorporate this bond. For example, if a chosen bond was going in +z direction: … -> (0,0,0) -> (0,0,1) -> … , and we attempted to grow it in the -y direction (chosen randomly out of +x, -x, +y, -y), we would check positions (0,-1,0) and (0,-1,1). If both of them were free, the polymer sequence would be changed to … -> (0,0,0) -> (0,-1,0) -> (0,-1,1) -> (0,0,1) -> … . If at least one of the positions of the shifted bond was occupied, selection of the random bond was repeated. The process was repeated until the polymer grew to the desired length. Since no polymer fibers can pass between the old bond and a kink, this process preserves the original topology and creates an un-entangled polymer.

**Calculating and analyzing heatmaps**
To obtain heatmaps, we first found all contacts within each polymer conformation. A contact was defined as two monomers being at a distance less than 2 monomer diameters. Contacts for all pairs of monomers were then put on a heatmap (i.e. a 2000-monomer polymer produced a 2000x2000 heatmap). When calculating contact frequency ratios for insulation and facilitation, averaging was performed over small regions of the heatmap to reduce sampling noise. Unless noted, we report the average value for insulation over a region of the heatmap, by averaging over monomers in the promoter-proximal third of the loop and over a +/-3 monomer E-P separation. The range of insulation values for different positions in the promoter-proximal half of the loop is shown in Figure **3B**. For facilitation, we average over a region of the heatmap defined by a +/– 3 monomer E-P distance and a +/– ((E-P genomic distance – loop length) / 6) monomer offset from a symmetric placement of an E-P pair around the loop bases (e.g. +/– 6 monomers in Figure **2B**). For the case when the E-P pair was within the loop, no averaging was used, since this occurs at short E-P distances where many contacts occur. To calculate the contact frequency ratio, we used simulations without a loop to calculate the expected frequency; for all parameter values, two simulations without a loop and ten simulations with a loop were performed, with a newly generated starting conformation for each simulation.

**Calculating spatial density around a loop**

To calculate normalized spatial density around the loop, we analyzed the model with default parameters (i.e. as in Figure **2**), where a loop connects monomers 970 and 1030 of a 2000-monomer ring.

1. For each conformation, we found the center of mass (COM) of the 60 loop monomers. We then defined "distal" monomers as monomers 0-950 and 1050-2000 (outside of the loop, plus 20 monomers, or 10kb, away from the loop base). We then counted how many distal monomers were at each spatial distance from the loop COM, averaging over all conformations and using bins starting at 0 with a step size of .5 monomers.

2. We then account for the fact that at larger spatial distances, a greater portion of the spatial shell exceeds the confining boundary. To this end, we took the position of the loop center of mass in a given conformation, and performed step 1 for a COM from a conformation in a different run (i.e., for conformation X in run Y, we took COM of the loop from the conformation X in the run (Y+1) modulus 10).

3. We then divided spatial densities from 1 by spatial densities from 2 to obtain normalized spatial densities.

To create a no-loop control, we repeated steps 1-3 for the monomers exactly opposite from the center of the loop, exactly repeating the same procedure (i.e. assuming 60-monomer loop began at monomer 1970, recalling that the polymer is closed into a length-2000 ring).

**Simulated FISH distributions**

To calculate simulated FISH distributions, we considered an E-P distance at which the magnitude of insulation and facilitation are comparable (90kb) and analyzed the model with default parameters. We then iterated over conformations and calculated the spatial distance between the E-P pair for both arrangements as well as for a control arrangement without a loop (i.e. monomers exactly opposite from the center of the loop, as for the spatial density calculation). The spatial distances were binned starting at 0 with a step size of 0.5 monomers.

**Acknowledgments**
We thank Mirny lab members for useful discussions.

# References

1. Bulger M, Groudine M (2011) Functional and mechanistic diversity of distal transcription enhancers. Cell 144: 327-339.
2. Levine M (2010) Transcriptional enhancers in animal development and evolution. Curr Biol 20: R754-763.
3. Pennacchio LA, Bickmore W, Dean A, Nobrega MA, Bejerano G (2013) Enhancers: five essential questions. Nat Rev Genet 14: 288-295.
4. Tolhius B, Palstra RJ, Splinter E, Grosveld F, de Laat W (2002) Looping and Interaction between Hypersensitive Sites in the Active B-globin Locus. Molecular Cell 10: 1453-1465.
5. Consortium EP, Bernstein BE, Birney E, Dunham I, Green ED, et al. (2012) An integrated encyclopedia of DNA elements in the human genome. Nature 489: 57-74.
6. Jin F, Li Y, Dixon JR, Selvaraj S, Ye Z, et al. (2013) A high-resolution map of the three-dimensional chromatin interactome in human cells. Nature 503: 290-294.
7. Andrey G, Montavon T, Mascrez B, Gonzalez F, Noordermeer D, et al. (2013) A Switch Between Topological Domains Underlies HoxD Genes Collinearity in Mouse Limbs. Science 340: 1195-1205.
8. Montavon T, Soshnikova N, Mascrez B, Joye E, Thevenet L, et al. (2011) A regulatory archipelago controls Hox genes transcription in digits. Cell 147: 1132-1145.
9. Dixon JR, Selvaraj S, Yue F, Kim A, Li Y, et al. (2012) Topological domains in mammalian genomes identified by analysis of chromatin interactions. Nature 485: 376-380.
10. Nora EP, Lajoie BR, Schulz EG, Giorgetti L, Okamoto I, et al. (2012) Spatial partitioning of the regulatory landscape of the X-inactivation centre. Nature 485: 381-385.
11. Sanyal A, Lajoie BR, Jain G, Dekker J (2012) The long-range interaction landscape of gene promoters. Nature 489: 109-113.
12. Phillips-Cremins JE, Sauria ME, Sanyal A, Gerasimova TI, Lajoie BR, et al. (2013) Architectural protein subclasses shape 3D organization of genomes during lineage commitment. Cell 153: 1281-1295.
13. Bondarenko VA, Jiang YI, Studitsky VM (2003) Rationally designed insulator-like elements can block enhancer action in vitro. EMBO Journal 22: 4728-4737.
14. Gohl D, Aoki T, Blanton J, Shanower G, Kappes G, et al. (2011) Mechanism of chromosomal boundary action: roadblock, sink, or loop? Genetics 187: 731-748.
15. Majumder P, Cai HN (2003) The functional analysis of insulator interactions in the Drosophila embryo. Proc Natl Acad Sci U S A 100: 5223-5228.
16. Raab JR, Kamakaka RT (2010) Insulators and promoters: closer than we think. Nat Rev Genet 11: 439-446.
17. Kulaeva OI, Nizovtseva EV, Polikanov YS, Ulianov SV, Studitsky VM (2012) Distant activation of transcription: mechanisms of enhancer action. Mol Cell Biol 32: 4892-4897.
18. Gerasimova TI, Corces VG (2001) Chromatin Insulators and Boundaries: Effects on Transcription and Nuclear Organization. Annual Review Genetics 35: 193-208.
19. Geyer PK, Clark I (2002) Protecting against promiscuity: the regulatory role of insulators. Cellular and Molecular Life Sciences 59: 2112-2127.
20. Savitskaya E, Melnikova L, Kostuchenko M, Kravchenko E, Pomerantseva E, et al. (2006) Study of long-distance functional interactions between Su(Hw) insulators that can regulate enhancer-promoter communication in Drosophila melanogaster. Mol Cell Biol 26: 754-761.

21. Kyrchanova O, Maksimenko O, Stakhov V, Ivlieva T, Parshikov A, et al. (2013) Effective blocking of the white enhancer requires cooperation between two main mechanisms suggested for the insulator function. PLoS Genet 9: e1003606.
22. Kurukuti S, Tiwari VK, Tavoosidana G, Pugacheva E, Murrell A, et al. (2006) CTCF binding at the H19 imprinting control region mediates maternally inherited higher-order chromatin conformation to restrict enhancer access to Igf2. Proc Natl Acad Sci U S A 103: 10684-10689.
23. Comet I, Schuettengruber B, Sexton T, Cavalli G (2010) A chromatin insulator driving three-dimensional Polycomb response element (PRE) contacts and Polycomb association with the chromatin fiber. Proc Natl Acad Sci U S A 108: 2294-2299.
24. Chetverina D, Aoki T, Erokhin M, Georgiev P, Schedl P (2014) Making connections: insulators organize eukaryotic chromosomes into independent cis-regulatory networks. Bioessays 36: 163-172.
25. Scott KS, Taubman, A.D., Geyer, P.K. (1999) Enhancer Blocking by the Drosophila gypsy Insulator Depends Upon Insulator Anatomy and Enhancer Strength. Genetics 153: 787-798.
26. Monod J, Changeux JP, Jacob F (1963) Allosteric Proteins and Cellular Control Systems. Journal of Molecular Biology 6: 306-329.
27. Wyman J (1964) Linked functions and reciprocal effects in hemoglobin: a second look. Adv Protein Chem 19: 91.
28. Mukhopadhyay S, Schedl P, Studitsky VM, Sengupta AM (2011) Theoretical analysis of the role of chromatin interactions in long-range action of enhancers and insulators. Proc Natl Acad Sci U S A 108: 19919-19924.
29. Giorgetti L, Galupa R, Nora EP, Piolot T, Lam F, et al. (2014) Predictive polymer modeling reveals coupled fluctuations in chromosome conformation and transcription. Cell 157: 950-963.
30. Joti Y, Hikima T, Nishino Y, Kamada F, Hihara S, et al. (2012) Chromosomes without a 30-nm chromatin fiber. Nucleus 3: 404-410.
31. Fussner E, Ching RW, Bazett-Jones DP (2011) Living without 30nm chromatin fibers. Trends Biochem Sci 36: 1-6.
32. Naumova N, Imakaev M, Fudenberg G, Zhan Y, Lajoie BR, et al. (2013) Organization of the mitotic chromosome. Science 342: 948-953.
33. Halverson JD, Smrek J, Kremer K, Grosberg AY (2014) From a melt of rings to chromosome territories: the role of topological constraints in genome folding. Rep Prog Phys 77: 022601.
34. Eastman P, Friedrichs MS, Chodera JD, Radmer RJ, Bruns CM, et al. (2013) OpenMM 4: A Reusable, Extensible, Hardware Independent Library for High Performance Molecular Simulation. J Chem Theory Comput 9: 461-469.
35. Dekker J, Marti-Renom MA, Mirny LA (2013) Exploring the three-dimensional organization of genomes: interpreting chromatin interaction data. Nat Rev Genet 14: 390-403.
36. Fudenberg G, Mirny LA (2012) Higher-order chromatin structure: bridging physics and biology. Curr Opin Genet Dev 22: 115-124.
37. Dorier J, Stasiak A (2009) Topological origins of chromosomal territories. Nucleic Acids Res 37: 6316-6322.
38. Rosa A, Everaers R (2008) Structure and dynamics of interphase chromosomes. PLoS Comput Biol 4: e1000153.

39. Benedetti F, Dorier J, Burnier Y, Stasiak A (2013) Models that include supercoiling of topological domains reproduce several known features of interphase chromosomes. Nucleic Acids Res 42: 2848-2855.
40. Le TB, Imakaev MV, Mirny LA, Laub MT (2013) High-resolution mapping of the spatial organization of a bacterial chromosome. Science 342: 731-734.
41. V. C. Littau VGA, J. H. Frenster, A. E. Mirsky (1964) Active and Inactive Regions of Nuclear Chromatin as Revealed by Electron Microscope Autoradiography. Proc Natl Acad Sci U S A 52: 93-100.
42. Loven J, Orlando DA, Sigova AA, Lin CY, Rahl PB, et al. (2012) Revisiting global gene expression analysis. Cell 151: 476-482.
43. Zhu X, Ling J, Zhang L, Pi W, Wu M, et al. (2007) A facilitated tracking and transcription mechanism of long-range enhancer function. Nucleic Acids Res 35: 5532-5544.
44. Harmston N, Lenhard B (2013) Chromatin and epigenetic features of long-range gene regulation. Nucleic Acids Res 41: 7185-7199.
45. Vendruscolo M (2011) Protein regulation: the statistical theory of allostery. Nat Chem Biol 7: 411-412.
46. Mirny LA (2010) Nucleosome-mediated cooperativity between transcription factors. Proc Natl Acad Sci U S A 107: 22534-22539.
47. Miller JA, Widom J (2003) Collaborative Competition Mechanism for Gene Activation In Vivo. Molecular and Cellular Biology 23: 1623-1632.
48. Li G, Levitus M, Bustamante C, Widom J (2005) Rapid spontaneous accessibility of nucleosomal DNA. Nat Struct Mol Biol 12: 46-53.
49. Belmont AS (2014) Large-scale chromatin organization: the good, the surprising, and the still perplexing. Curr Opin Cell Biol 26: 69-78.
50. Nagano T, Lubling Y, Stevens TJ, Schoenfelder S, Yaffe E, et al. (2013) Single-cell Hi-C reveals cell-to-cell variability in chromosome structure. Nature 502: 59-64.

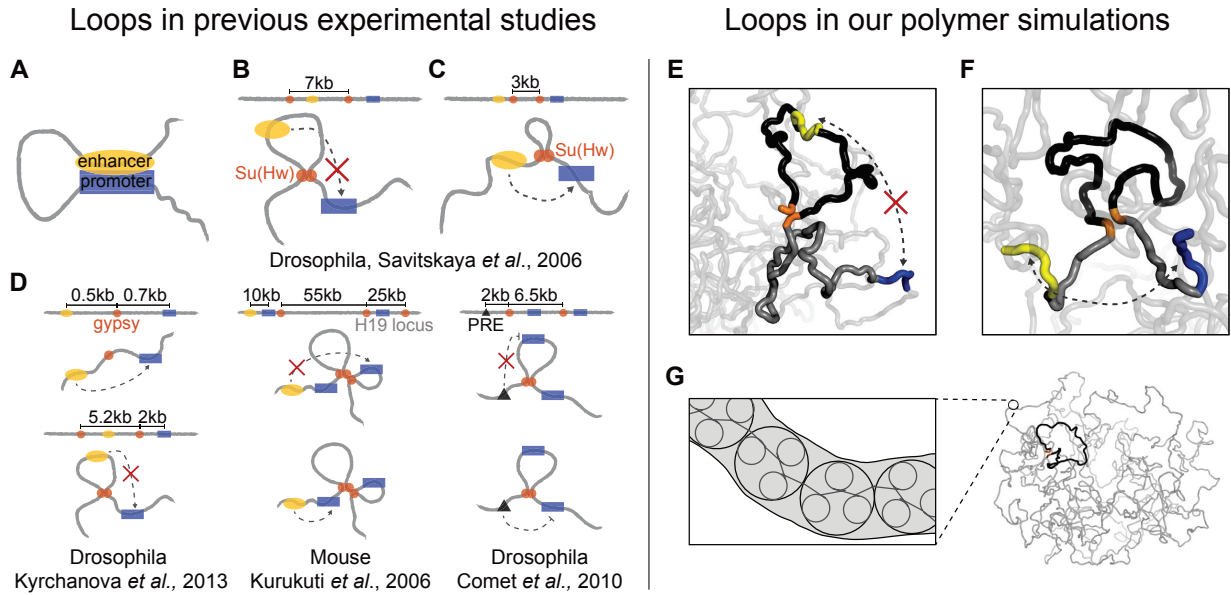

**Figure 1. Enhancer-promoter pairs in the context of other interactions.** *Experimental Studies*, (**A**) Illustration of an enhancer (in yellow) spatially interacting with a promoter (blue) along a chromatin fiber. This coloring convention continues throughout the paper. (**B**) A recent study in Drosophila suggested a 7kb chromatin loop formed between Su(Hw) insulators (orange) could decrease E-P interactions (red "X") [20]. (**C**) Conversely, a 3kb chromatin loop in the region between enhancer and promoter was proposed to increase E-P interactions. (**D**) Five arrangements for proposed looping interactions from three studies, left to right, [21], [22], and [23]. (*left*) a single Drosophila *gypsy* element between an enhancer and a promoter did not change their interactions (*top*), however an additional *gypsy* element upstream of the enhancer decreased E-P interactions (*bottom*) [21]. (*center*) at the mouse H19 locus, a regulatory element with multiple larger loops (55kb and 25kb) was suggested to control multiple E-P contacts; the enhancer can regulate the promoter before the loop, but cannot regulate the promoter within the loop [22]. (*right*) chromatin loops may also modulate spatial interactions between silencing elements (e.g. PRE, black triangles) and their target promoters [23]. The promoter within the loop is not silenced (*top*), whereas the promoter beyond the loop is silenced (*bottom*). *Polymer Simulations*, (**E**) Arrangement 1: polymer conformation where an enhancer is within a chromatin loop and a promoter is beyond the loop. (**F**) Arrangement 2: polymer conformation where an enhancer is before the loop and a promoter is after the loop. (**G**) (*left*) zoom-in on our polymer model of chromatin. The three large circles represent one monomer each; each monomer consists of three nucleosomes (small circles) or 500bp. (*right*) full view of a sample polymer conformation showing a 30kb chromatin loop (black) with highlighted loop-bases (orange) within a 1Mb region.

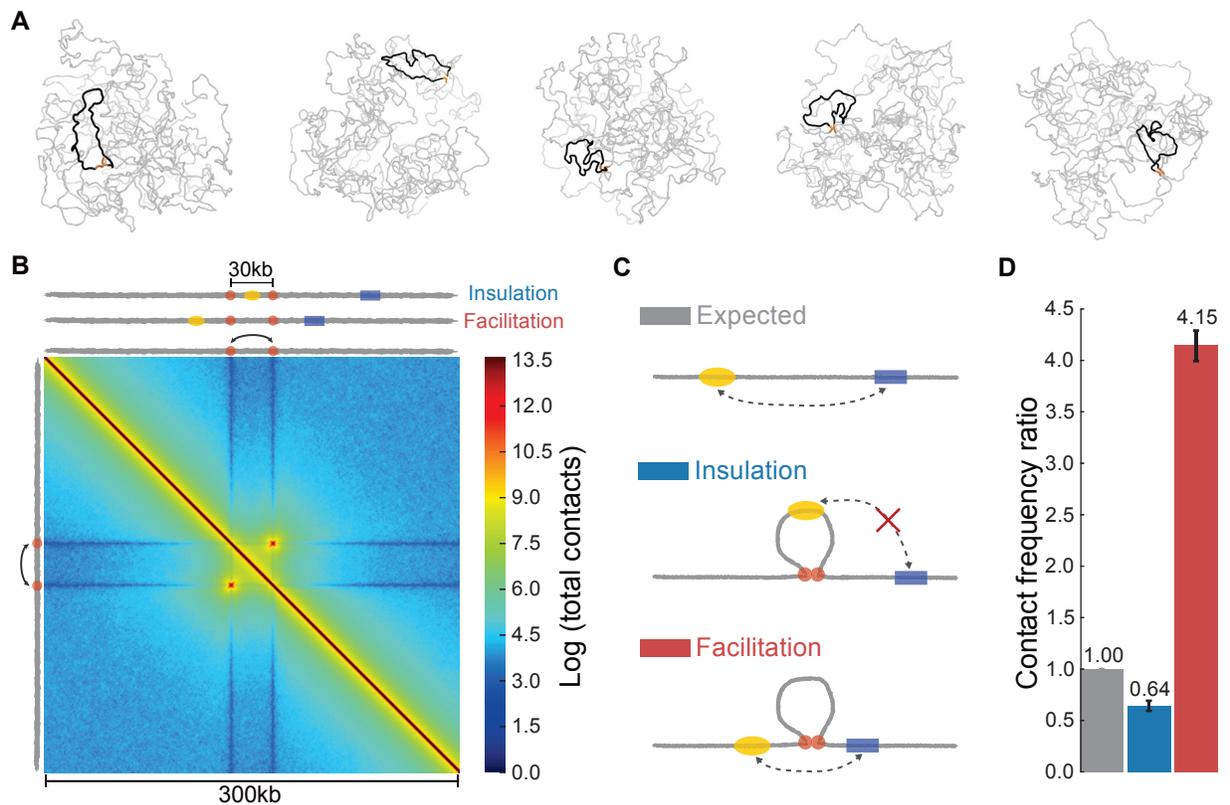

**Figure 2. A chromatin loop alters the frequency of enhancer-promoter interactions.** (**A**) Five sample conformations from polymer simulations with a 30kb permanent loop (black) formed between two loop bases (orange) in a 1Mb region of fiber. (**B**) Average heatmap (300kb by 300kb) for polymer simulations of the permanent, one-loop system, with a 30kb loop (aggregated over 800,000 simulated conformations). Top and left edges show positions of the enhancer (yellow), promoter (blue), and loop bases (orange) for insulation and facilitation arrangements. (**C**) Schematics of E-P arrangements. (*top*) chromatin fiber without a fixed loop and with E-P genomic distance of 50kb, as used to calculate expected (no-loop) contact frequencies (Methods). (*middle*) arrangement where insulation is observed, represented by the red "X". (*bottom*) arrangement where facilitation is observed. (**D**) Contact frequency ratios (Methods) for insulation and facilitation arrangements with a 30kb loop and 50kb E-P genomic distance. Here and below, error bars indicate one standard deviation about the mean.

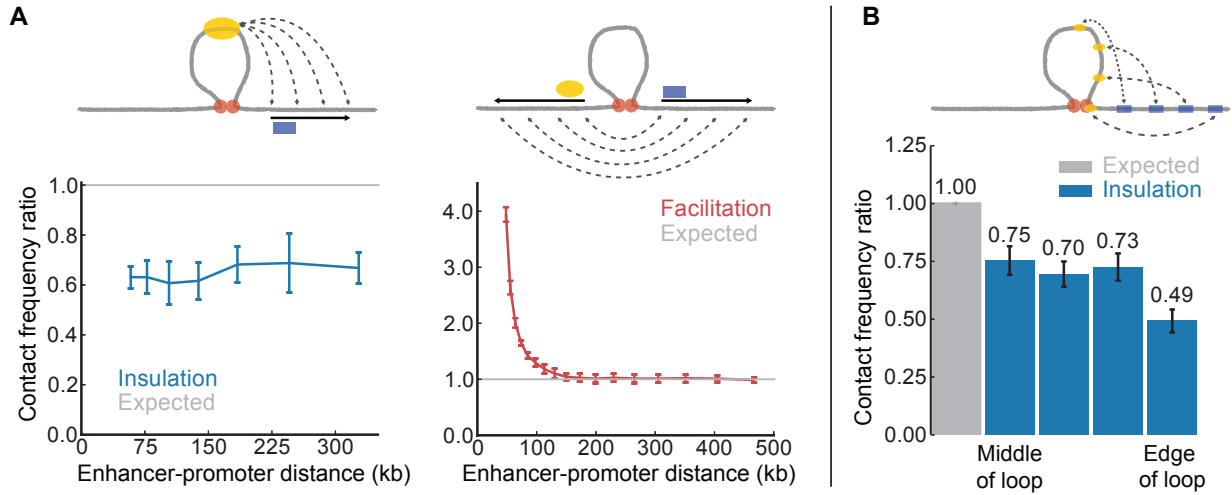

**Figure 3. Insulation and facilitation strength depends on enhancer-promoter positions.** (**A**) Insulation (*left*) and facilitation (*right*) as a function of E-P genomic distance. For insulation, enhancer position remains fixed. For facilitation, an E-P pair is positioned symmetrically around the loop at each genomic distance. (**B**) Insulation for different positions of the enhancer within the loop with a constant genomic distance of 50kb.

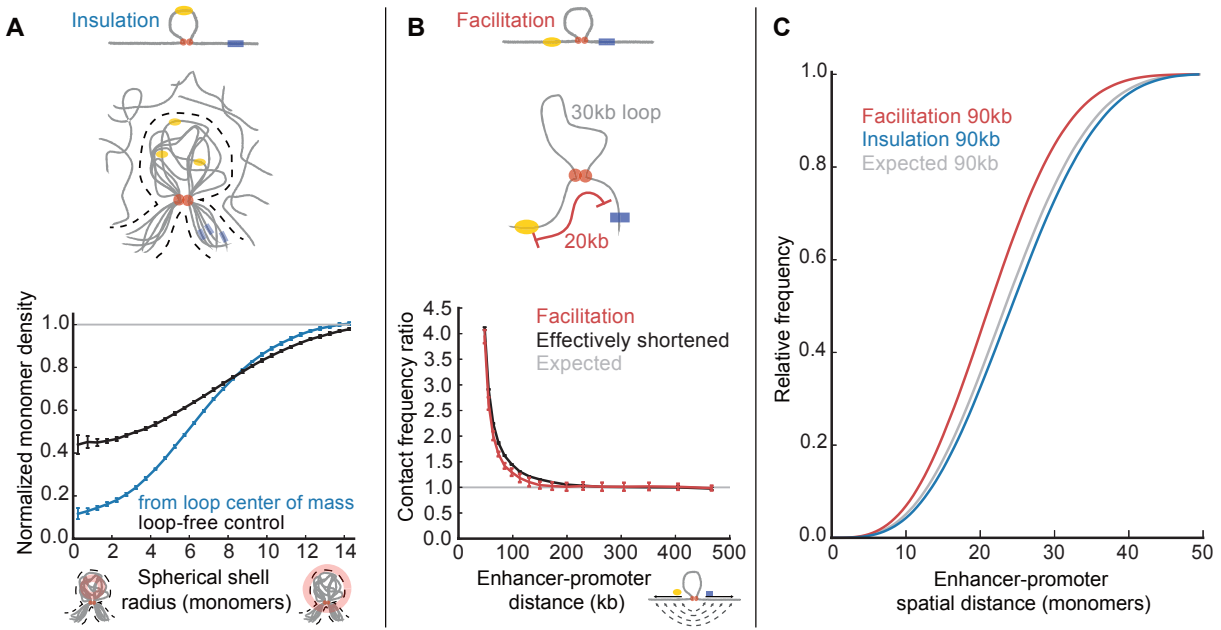

**Figure 4. Mechanisms of insulation and facilitation.** (**A**) (*top*) Illustration of the insulation mechanism: strong dynamic steric exclusion by a chromatin loop is shown by a superposition of loops in multiple conformations (grey, with enhancer and promoter) and their sterically excluded region (dashed lines), surrounded by other distal regions of chromatin (grey). (*bottom*) Density of distal monomers (i.e. outside the loop and >10kb from the loop base) as a function of radial distance from the center of mass of the loop. The loop-free control exactly repeats this procedure for an equivalent region without a loop. Both are normalized using respective radial-position dependent spatial density (Methods). (**B**) (*top*) Illustration of facilitation mechanism: an E-P pair flanking a loop has an effectively shorter genomic distance; here an E-P pair with 50kb separation and a 30kb loop behaves similarly to an E-P pair separated by 20kb in a region without a loop. (*bottom*) Comparison of contact frequency ratios for the above situations, as a function of E-P distance. (**C**) Simulated cumulative distribution of spatial distances (*in silico* FISH) for an E-P pair with a genomic distance of 90kb.

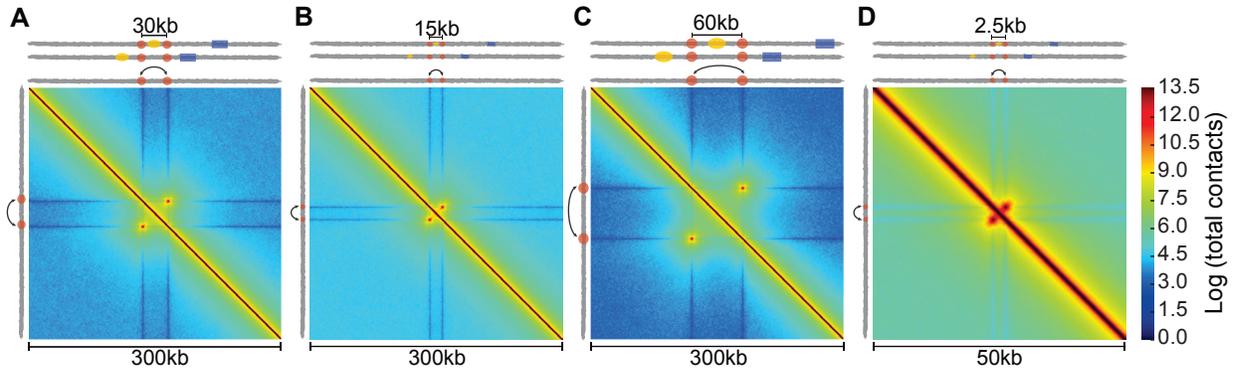

**Figure S1: Effects of loop size.** Schematics show insulation and facilitation arrangements including the enhancer (yellow), the promoter (blue), and the loop bases (orange) for the heatmaps below. In all cases the main qualitative features remain the same. (**A**) A 300kb by 300kb heatmap for a 30kb loop, as shown in Figure 2B. (**B**) A 300kb by 300kb heatmap for a smaller loop of length 15kb. (**C**) A 300kb by 300kb heatmap for a larger loop of length 60kb. (**D**) A 50kb by 50kb heatmap for a very small loop of length 2.5kb. In this simulation only, each monomer represents 250bp rather than 500bp of a more flexible fiber (k=2, see Methods), representing a loosely arranged chromatin fiber. This heatmap indicates that insulation and facilitation may still manifest at smaller genomic distances for a more flexible or loosely packed chromatin fiber, as these changes cause small loops to behave similarly to larger loops. Note the color of the map differs due to the smaller dynamic range in total number of interactions for this shorter chromatin fiber, but the same qualitative features are present.

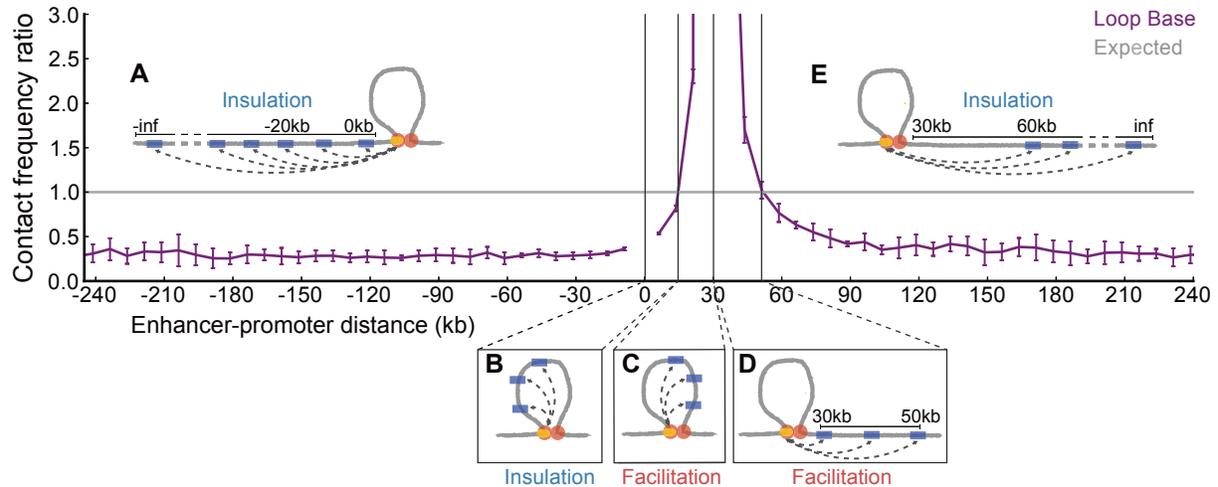

**Figure S2: Loop-base profile.** Contact frequency ratio of the loop base vs. all other loci (i.e. a 4C-like profile); an enhancer placed at one loop base (0kb) displays a complex pattern of insulation and facilitation, which we summarize in terms of five regions (**A-E**). The x-axis shows the upstream or downstream distance to the loop base where this enhancer is placed; note the position of the other loop base is at 30kb. The y-axis is truncated at contact frequency ratios of 3.0, as when both the enhancer and promoter are positioned at loop bases (i.e. x=30kb), the magnitude of facilitation is very large since the loop bases are always in contact. (**A**) Insulation of the loop base from upstream regions of chromatin. (**B**) Intra-loop insulation when E-P distance is less than half the loop size. (**C**) Intra-loop facilitation when E-P distance exceeds half the loop size. (**D**) Facilitation when the E-P distance slightly exceeds the loop size. (**E**) Insulation of the loop base from distal downstream regions of chromatin.

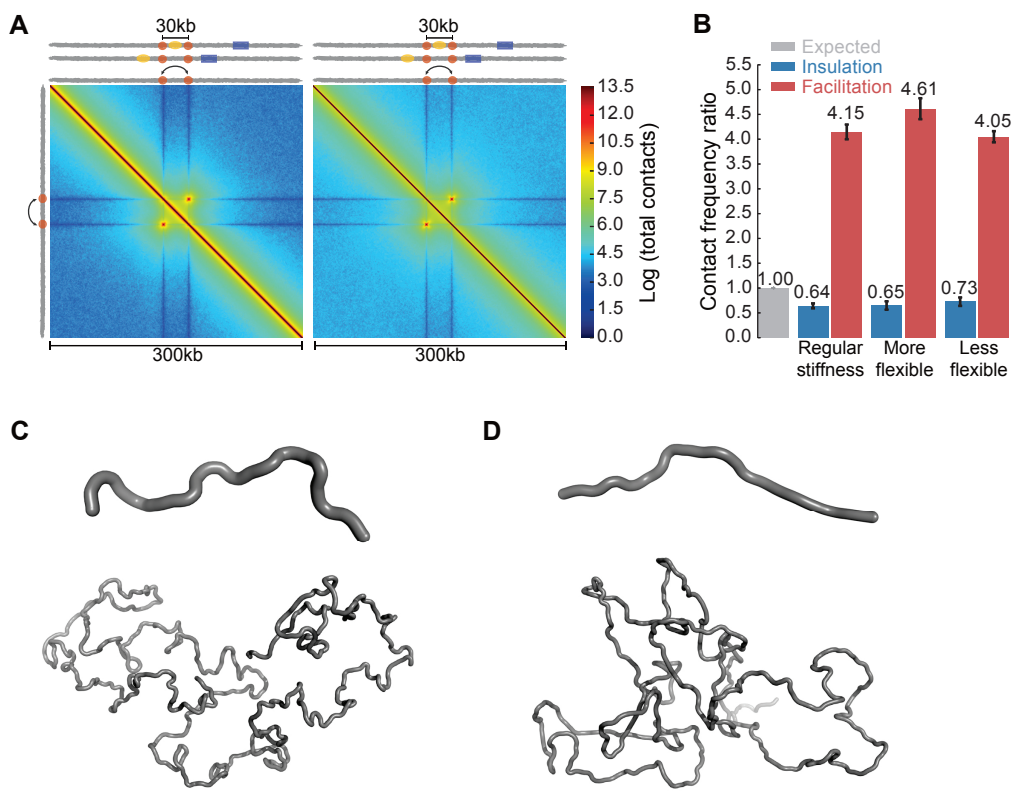

**Figure S3: Effects of chromatin fiber flexibility.** (**A**) Heatmap on left displays log (total # of contacts) for simulations with a more flexible polymer and standard parameters: 30kb chromatin loop, 2% density, fiber crossing (topoisomerase activity). On the right is a heatmap for the less flexible polymer. In both cases, the loop features observed in Figure **2B** are still present. (**B**) Bar plot shows insulation and facilitation: at the stiffness presented in the main figures, for a more flexible polymer, and for a less flexible polymer. (**C**) (*top*) shows a 20 monomer or 10kb stretch from a conformation of a more flexible polymer. (*bottom*) shows a 500 monomer or 250kb region from a conformation of a more flexible polymer. (**D**) Same as (C), but for a less flexible polymer. Note the smoother appearance of the less flexible chromatin fiber conformation.

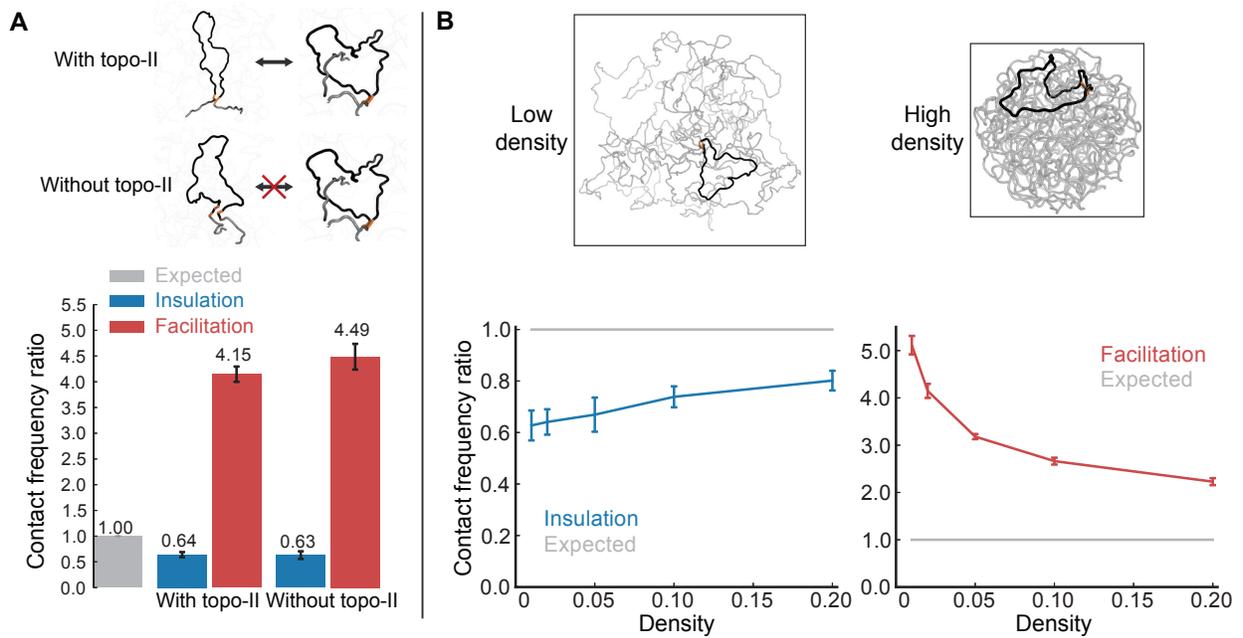

**Figure S4: Effect of topoisomerase and chromatin density on local loop-mediated interactions.** (**A**) Effect of topological constraints on insulation and facilitation. With topo-II, there are no topological constraints and a conformation without chromatin threaded through the loop can convert to a conformation with chromatin threaded through the loop. Without topo-II (with topological constraints), chromatin fibers cannot cross and the two conformations cannot interconvert. Bar plot shows the contact frequency ratio for an E-P genomic distance of 50kb. (**B**) Effect of density on insulation and facilitation; bar plots show results for 50kb E-P genomic distance.

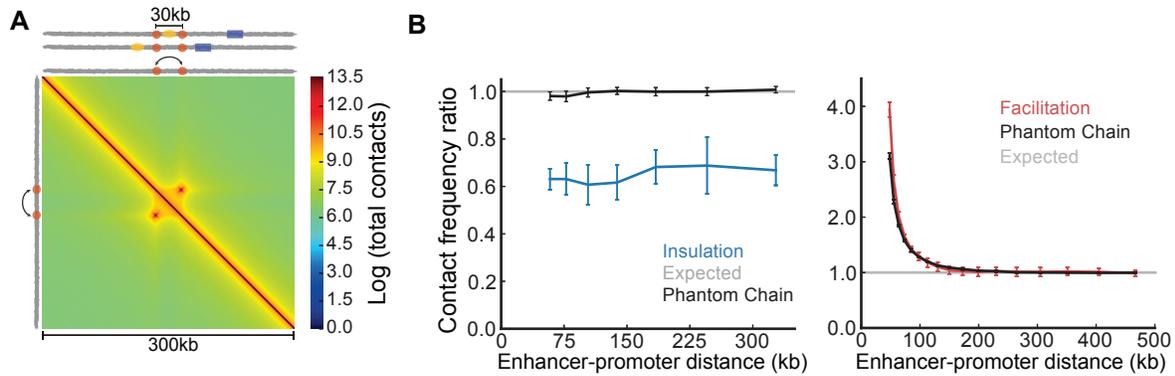

**Figure S5: Effects of phantom polymer chain.** (**A**) Heatmap for phantom polymer chain with a 30kb loop, where insulation and facilitation arrangements are shown as in Figure **2B**. The vertical and horizontal stripes of depleted interactions are almost non-existent, indicating dramatically reduced insulation. (**B**) Bar plot displays insulation and facilitation for the regular scenario (Figure **2B**) on the left and the phantom chain on the right. Facilitation is slightly diminished, whereas insulation completely disappears.

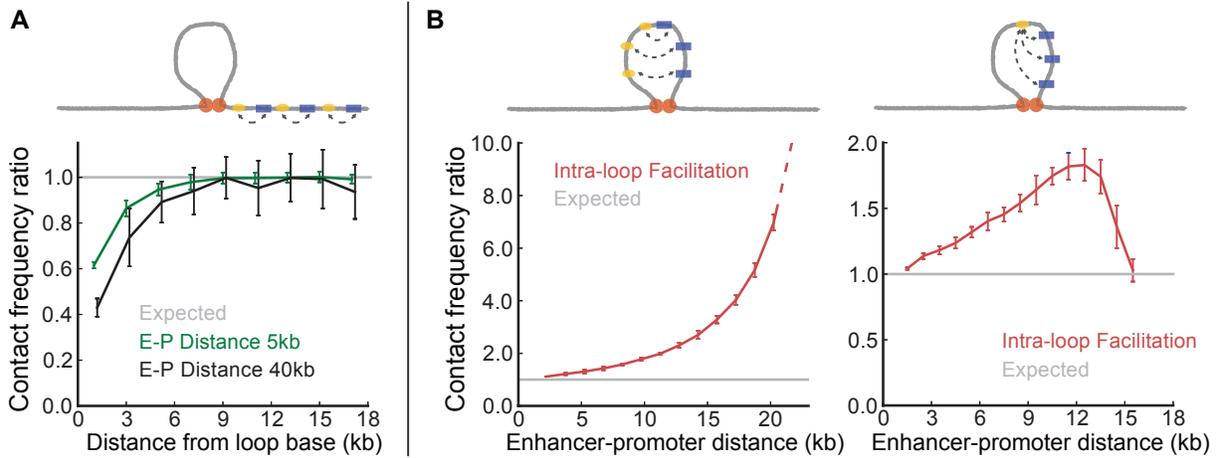

**Figure S6: Loop shadowing and intra-loop facilitation** (**A**) Regions immediately outside the loop are also sterically excluded by the loop; in other words the loop's steric "shadow" can cause insulation when the E-P pair is near, but outside, of the loop. Note the black line has been slightly offset, so that error bars are visible. (**B**) (*left*) Intra-loop facilitation when the E-P pair is positioned symmetrically within the loop. When E-P distances are much less than the loop size, the loop has a negligible influence on their contact frequency, and the contact frequency ratio is ~1. However the magnitude of facilitation increases very quickly as E-P distance approaches the loop size because the loop bases are always in contact (corresponding strong peak in Figure **S2** at 30kb = loop size). Note truncated y-axis (at contact frequency ratios of 10.0). (*right*) Asymmetric E-P placement with increasing E-P distance, where the enhancer stays in the middle of the loop, while the promoter moves towards the loop base. Intra-loop facilitation drops off approaching 15kb (half of the loop size), due to a superposition with the insulating properties of the loop bases.

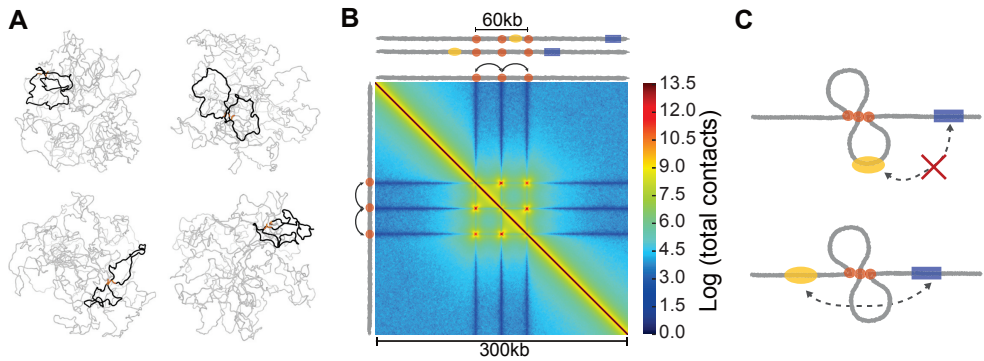

**Figure S7: Effects of two consecutive loops.** (**A**) Four sample polymer conformations from simulations of the two-loop system with loops (black) and loop bases (orange) highlighted. (**B**) Heatmap which shows log (total # of contacts) for the two-loop system. Each loop is 30kb in the 300kb by 300kb region shown. The four red dots closer to the diagonal are the direct interaction of the loop bases from the formation of two loops. The two, weaker, red dots further from the diagonal are the interaction between the base at the start of the first loop and at the end of the second loop. The horizontal and vertical stripes of darker blue are indicative of strong insulation. Annotations show two loops formed from three bases (orange) along with the insulation and facilitation E-P placements. (**C**) Schematics of E-P arrangement for the two-loop system (*top*) insulation, indicated the red "X", (*bottom*) facilitation.

| Name | Figure # | Fiber crossing | Density | Flexibility (k) | Loop size (kb) | Excluded volume | # of loops | 1 monomer = x bps |
|---|---|---|---|---|---|---|---|---|
| Standard | 2,3,4,S2,S6 | yes | 0.02 | 3 | 30 | yes | 1 | 500 |
| Without topo-II | S4 | no | 0.02 | 3 | 30 | yes | 1 | 500 |
| Low density | S4 | yes | 0.01 | 3 | 30 | yes | 1 | 500 |
| Density 0.05 | S4 | yes | 0.05 | 3 | 30 | yes | 1 | 500 |
| Density 0.1 | S4 | yes | 0.1 | 3 | 30 | yes | 1 | 500 |
| High density | S4 | yes | 0.2 | 3 | 30 | yes | 1 | 500 |
| More flexible | S3 | yes | 0.02 | 2 | 30 | yes | 1 | 500 |
| Less flexible | S3 | yes | 0.02 | 4 | 30 | yes | 1 | 500 |
| Smaller loop | S1 | yes | 0.02 | 3 | 15 | yes | 1 | 500 |
| Larger loop | S1 | yes | 0.02 | 3 | 60 | yes | 1 | 500 |
| Smallest loop | S1 | yes | 0.02 | 2 | 2.5 | yes | 1 | 250 |
| Phanton chain | S5 | yes (transparent chain) | 0.02 | 3 | 30 | no | 1 | 500 |
| Two-loop | S7 | yes | 0.02 | 3 | 30 per loop | yes | 2 | 500 |

**Table S1:** List of parameter values for all presented simulations.

**Video S1:** Langevin Dynamics of a 30kb permanent loop formed in a 1Mb region of chromatin fiber, as in Figure 2. The polymer is colored according to the facilitation arrangement, where the loop (black) occurs in the region between the enhancer (yellow) and the promoter (blue). The movie is presented at a rate of 1000 simulated time-steps per one second of real time; every seventh frame of the movie corresponds to a computationally obtained conformation, with quadratic interpolation performed between subsequent conformations (done to avoid high-frequency fluctuations between neighboring frames). The total simulation time for each run of each parameter set was approximately 5,000 times longer than the part of simulation displayed in this movie.